\documentclass[12pt]{iopart}

\usepackage[backend=biber,style=science,]{biblatex}
\usepackage[titletoc, title, page]{appendix}
\usepackage{graphicx}
\usepackage{multirow}
\usepackage{hyperref}
\usepackage{lscape}
\usepackage{tikz}
\usepackage{subcaption}
\usetikzlibrary{shapes.geometric, arrows}

\bibliography{References}

\tikzstyle{Topic} = [rectangle, rounded corners, 
minimum width=3cm, 
minimum height=1cm,
text centered, 
draw=black, 
fill=red!30]

\tikzstyle{Intervention} = [trapezium, 
trapezium stretches=true, 
trapezium left angle=70, 
trapezium right angle=110, 
minimum width=3cm, 
minimum height=1cm, text centered, 
draw=black, fill=blue!30]

\tikzstyle{Details} = [rectangle, 
minimum width=3cm, 
minimum height=1cm, 
text centered, 
text width=3cm, 
draw=black, 
fill=orange!30]

\tikzstyle{decision} = [diamond, 
minimum width=3cm, 
minimum height=1cm, 
text centered, 
draw=black, 
fill=green!30]
\tikzstyle{arrow} = [thick,->,>=stealth]
 
\begin{document}

\title{Einsteinian gravitational concepts throughout secondary school}

\author{Corey McInerney and Phil Sutton}
\address{University of Lincoln, School of Mathematics and Physics; Lincoln, LN6 7TS, UK}
\ead{cmcinerney@lincoln.ac.uk}

\vspace{10pt}
\begin{indented}
\item[]\today                                                                                                          
\end{indented}

\begin{abstract}
Einstein's theory of relativity is largely thought of as one of the most important discoveries of the 20\textsuperscript{th} century and continues to pass observational tests over 100 years later. Yet, it is Newtonian gravity, a 350 year old formalism proven to be less accurate than relativity, which is taught in schools. It has been shown that Einsteinian gravitational concepts can be well understood by students in both primary and secondary education. In this paper, a cross-section of students from Yr 7-13 enrolled in an English secondary school took part in an intervention designed to introduce the idea of gravity from spacetime curvature. The overall aim of this work is to assess the viability of including relativity in the secondary curriculum and to ascertain which year this material would be best placed in. We determine that all year groups where able to appreciate the effects of curvature to some extent. Visual demonstrations aided conceptual understanding at Yr 7-8 level, but this does not have a strong effect on their ideas around the source of the gravitational force. Participants in Yr 9-13 were able to understand concepts beyond those introduced in the demonstrations. However, a deeper understanding of curvature as the source of the gravitational force is not seen until years 12 \& 13. We find that those in Yr 13 have the best overall understanding of the concepts introduced during our intervention.  
\end{abstract}

\vspace{2pc}
\maketitle

\section{Introduction}
\label{Introduction}

In 1687, Isaac Newton published his law of universal gravitation\textsuperscript{\cite{Newton}}, it is commonly taught in schools because it combines the concept of mass and weight with everyday phenomenon like falling objects\textsuperscript{\cite{Elise2007}}. Gravity is regarded as being one of the key threshold concepts in education used as an indicator of a students understanding of wider physics topics\textsuperscript{\cite{MeLa2005, FeLe2021}}. However, research suggests that a large number of misconceptions about gravity are held by students. These include issues surrounding the direction of gravitational force, gravity outside of the Earth and the role of the Sun in celestial dynamics\textsuperscript{\cite{Syuhendri2019, SmTr1988, PiBaTr1988}}. These issues are tied to students not being able to fully grasp where the gravitational force comes from, with some students believing that objects only have a gravitational pull if they are heavy and spherical, or that air is needed in order for gravity to act between objects\textsuperscript{\cite{GaBa1997, RuCaDu1985}}. Misconceptions can stay with students throughout their schooling, impacting confidence, ability and thus students overall enjoyment of science\textsuperscript{\cite{Driver1981}}.

A new approach pioneered by the Einstein-First project could eliminate some of the conceptual issues faced by students. The idea is to introduce principals from modern physics, such as Einstein's Theory of General Relativity (GR) into the Australian curriculum at primary and secondary level. A similar project, ReleQuant, in Norway also pioneers the teaching of Relativity and quantum mechanics at secondary school\textsuperscript{\cite{BuHeAn2015, KeHeVe2018}}.

Einstein's Theory of General Relativity provides a mathematical description of gravity and offers a qualitative description of how gravity works\textsuperscript{\cite{Moore}}. A key feature of GR is the non-Euclidean geometry of space and the interplay between time and space. The sophisticated mathematics involved in GR are seen as a barrier to students below degree level. The ReleQuant project bypasses this mathematical rigor by placing conceptual language at the forefront of the learning process, exploring concepts via thought experiments and visual demonstrations.

This study explores the impact of a one-off intervention on students' understanding of Einsteinian gravitational concepts. Similar studies have been performed in  Lebanon\textsuperscript{\cite{HaPe1972}}, Australia\textsuperscript{\cite{PiVeBl2014, KaBlMo2017b}}, Indonesia\textsuperscript{\cite{DuBlKa2020}} and Italy\textsuperscript{\cite{RuMaLe2021, Ruggiero2022}}. Each of those works report that students have a better understanding of gravity from a GR stand point after being introduced to concepts such as curvature, time dilation and spacetime diagrams. 

The intervention used for this study was designed to fit into a standard 50 minute lesson period at a single English state funded secondary school. Within that period, students were taught about how spacetime curvature produces the gravitational force with the help of two hands on activities. The host school is a fully government funded, selective, mixed grammar school located the town of Gainsborough, County Lincolnshire, England.
This study took place during February and March 2023 and at that time a total of 1237 students were enrolled at the school. 

An identical intervention was delivered to all age groups (years 7 - 13) at this school. Students were issued with questionnaires before and after the intervention and by comparing the results across year groups, we hope to identify the optimum place for GR to be introduced into the secondary science curriculum. 

In the following section we describe the research methodology in comparison with other works, as well as the content and application method of our intervention. In Section \ref{Results&Discussion}, we present the results of our research, which are also discussed and analysed.

\section{Method}
\label{Method}

\subsection{The Host School}
\label{Host}

The school where this research was performed is a fully government funded, selective, mixed grammar school located in England. 
At the time of this study, a total of 1237 students were enrolled at the school. The UK government website (\url{https://www.find-school-performance-data.service.gov.uk}) shows that in 2022, the school was ranked in the top 25\% in the UK for academic performance at GCSE level and that 82\% of pupils at the school achieved A-Level grades between A* and C.

\subsection{Comparison With Other Works}
\label{Comparisson}

In other works, interventions range from short, one off workshops\textsuperscript{\cite{ChFoKa2018}}, to full 20 session programs\textsuperscript{\cite{KaBlBu2018, KaBlSt2020}}. The offering a longer syllabus means more content and concepts can be covered, students can learn at their own pace and there are opportunities for deeper learning experiences. Nonetheless, it has been shown that one-off interventions have a significant impact on students engagement with science topics and consequently, their future career choices\textsuperscript{\cite{WyHa2009, LaLiTh2007}}. 

The intervention used in this paper was designed as a single, one-off session to introduce students to the subject of GR and to explore gravitational phenomena in the realm of GR. While other works have used participants from a single class, a single year group or a particular Key Stage (KS), we offered this intervention to all physics students in the school. A total of 183 students from years 7-13, choose to take part in this study, the breakdown of which is shown in Table \ref{Participants}.

\begin{table}[!ht]
\caption{Number of participants in this study by year group.}
\label{Participants}
\centering
\begin{tabular}{|cc|c|}
\hline
\multicolumn{1}{|c|}{\textbf{Key Stage}}     & \textbf{Year Group} & \textbf{Number of Students} \\ \hline
\multicolumn{1}{|c|}{\multirow{3}{*}{KS3}} & 7  & 59 \\ \cline{2-3}
\multicolumn{1}{|c|}{}    & 8  & 41 \\ \cline{2-3}
\multicolumn{1}{|c|}{}    & 9  & 24 \\ \hline
\multicolumn{1}{|c|}{\multirow{2}{*}{KS4}}                     & 10 & 20 \\ \cline{2-3}
\multicolumn{1}{|c|}{}                                           & 11 & 13 \\ \hline
\multicolumn{1}{|c|}{\multirow{2}{*}{A-Level}}                     & 12 & 12 \\ \cline{2-3}
\multicolumn{1}{|c|}{}                                           & 13 & 14 \\ \hline
\multicolumn{2}{|c|}{\textbf{Total}}                     & 183 \\
\hline
\end{tabular}
\end{table}

Surveying students from one single school allows us to see how students understanding of gravity changes throughout the educational years via this cross-sectional study and offers the opportunity for longitudinal studies in the future following specific cohort(s). Gravity is taught in England at KS3 (11-14 years old), KS4 (14-16 years old) and A-Level (16-18 years old), therefore, any misconceptions picked-up at KS3 have the potential to carry forward and impact a students understanding at A-Level.

\subsection{The Intervention}
\label{The Interventions}

The intervention implemented in this work consisted of 33 identical 50 minute workshops in the presence of 8 - 30 students from a single year group. The same researcher conducted all sessions, covering the same material and activities.

The intervention began with students being asked to describe gravity in their own words before GR was introduced. Table \ref{Table:Outline} shows the order that topics were introduced, the material spoken about and where any hands-on activities involved tie-in (highlighted in blue).

\begin{table}[h]
\caption{An outline of the different topics and concepts covered during our intervention.}
\label{Table:Outline}
\centerline{%
\begin{tabular}{|lp{10.5cm}|}
\hline
\multicolumn{1}{|c|}{\textbf{Topic}}                        & \multicolumn{1}{c|}{\textbf{Details}}                                                                    \\ \hline
\multicolumn{1}{|l|}{Newton's Law of Universal Gravitation} & Every particle attracts every other particle in the universe.                                            \\ \hline
\multicolumn{1}{|l|}{Gravitational field lines}             & Field lines on the surface of and around celestial bodies.                                               \\ \hline
\multicolumn{1}{|l|}{$E=mc^2$}                                 & Relation between mass and energy in Special Relativity, leading to relation between mass   and gravity in GR.            \\ \hline
\multicolumn{1}{|l|}{General Relativity}                    & Space, time and spacetime curvature.                                                                     \\ \hline
\multicolumn{2}{|c|}{\textcolor{blue}{\textbf{Spacetime simulator}}}                                                                                                                              \\ \hline
\multicolumn{1}{|l|}{Gravitational force in GR}             & Force increase with curvature.                                                                           \\ \hline
\multicolumn{2}{|c|}{\textcolor{blue}{\textbf{Triangles on balloons}}}                                                                                                                          \\ \hline
\multicolumn{1}{|l|}{Geometry in GR}                        & GPS corrections for Earth's curvature. Positive and negative curvature.   The curvature of the universe. \\ \hline
\end{tabular}%
}
\end{table}

\subsubsection{The Spacetime Simulator}
\label{SpacetimeSimulator}

A common analogy for the for curvature of spacetime uses a sheet of stretchy, elastic material such as Lycra, and some different massed objects. This particular analogy is commonly referred to as `the spacetime simulator'\textsuperscript{\cite{WhMoSl1993}} and has proven to be effective at dispelling misconceptions around gravity\textsuperscript{\cite{PoDe2021}}.

\begin{figure}[h]
	\centering
	\includegraphics[width=0.5\textwidth,height=0.5\textheight,keepaspectratio]{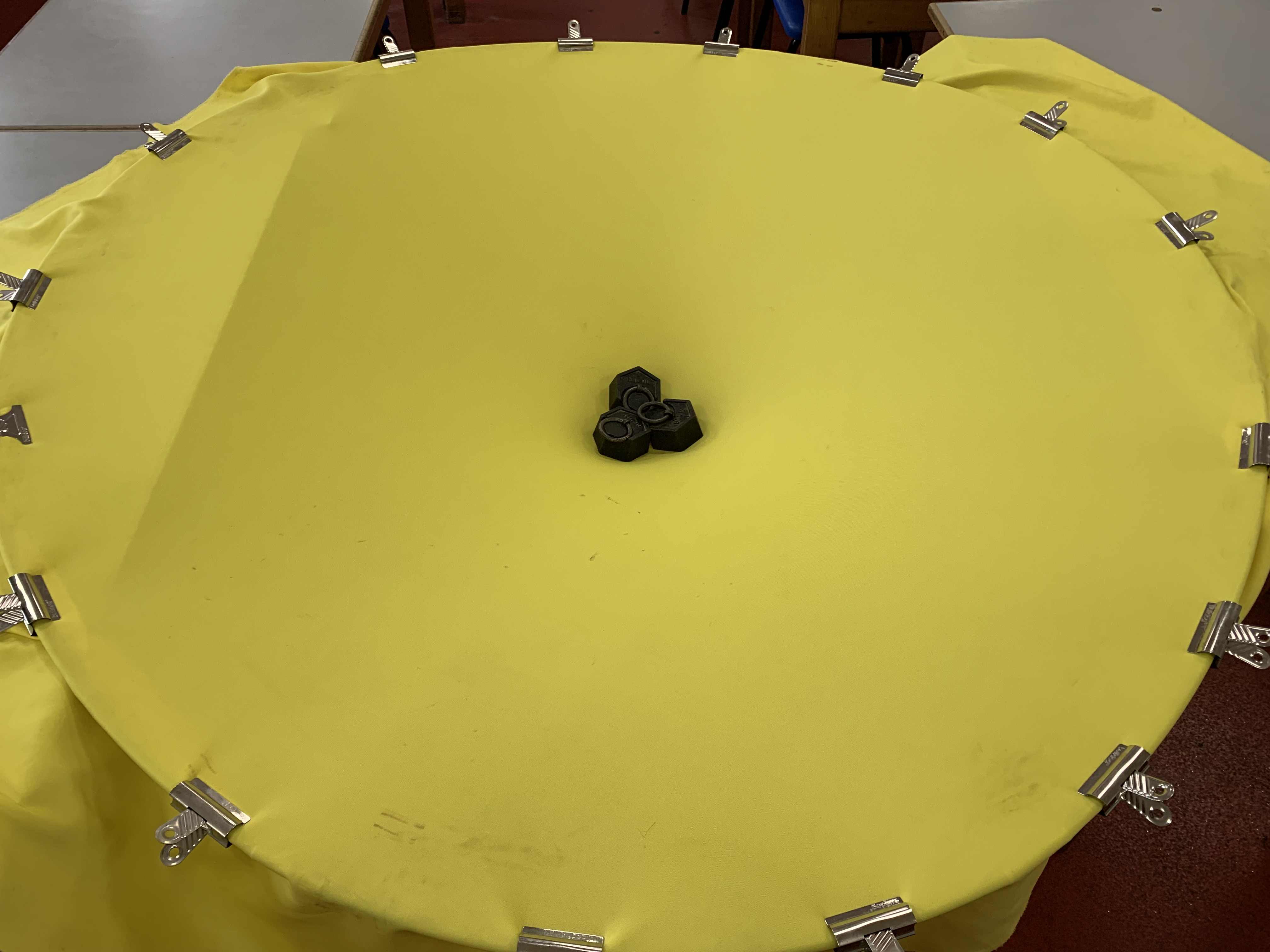}
	\caption{The stretched sheet of Lycra used as a spacetime simulator. Three 1 kg masses placed in the centre of the sheet provide curvature.}
	\label{fig:Sheet}
\end{figure}

For our demonstration, a large Lycra sheet was stretched flat across a plastic frame as seen in Figure \ref{fig:Sheet}. Three students were then invited forward to hold the sheet in the air and it was explained to the group that the sheet represents spacetime in the flat Euclidean plane (as per the Newtonian model). Students were then invited forward to roll tennis, squash, and golf balls across it, perceiving that they travel in straight lines and that mass does not affect motion in flat space. A mass heavy enough to deform the sheet was then placed on it and students observed that the balls now roll towards the central mass much like a ball rolling towards the bottom of a hill. By adding more mass to the centre of the sheet and increasing the curvature, students observe that the balls roll towards the central mass much quicker than before. This opens up a discussion around mass, curvature and gravitational attraction.

Additionally, while the sheet is deformed, squash balls were given some sideways velocity. Here, students observe the ball circling around the central mass with motion analogous to an orbit. It was important here to draw attention to the fact that the orbital motion decays due to the loss of energy from friction between the ball and the sheet, whereas orbits in the universe conserve energy in general and will be continuous. 

To explore this further, students were shown a plot akin to Figure 1b from Kaur \textit{et al.} (2017)\textsuperscript{\cite{KaBlMo2017a}} which illustrates that increasing curvature (adding mass to the sheet) changes the distance between two points in spacetime, thus influencing the geometry of the surrounding spacetime.

\subsubsection{The Geometry of General Relativity}
\label{Geometry}

Following the introduction to curvature, the intervention moved to exploring the geometry of spacetime.

Pairs of students were issued with different sized balloons. Each balloon had a triangle drawn on it and the students were tasked with carefully measuring the internal angles of the triangles by placing protractors on the balloons surface. Students then compared their results to the 180$^{\circ}$ of a triangle in Euclidean space to that of one with increased curvature and how that changes the internal angles. These results were then related to applications of non-Euclidean geometry such as the spherical curvature of Earth, it's effect on lines of longitude (such as their convergence at the poles), GR corrections to GPS and the potential geometries of the universe (hyperbolic, flat, parabolic).

\begin{figure}[h]
	\centering
	\includegraphics[width=0.5\textwidth,height=0.5\textheight,keepaspectratio]{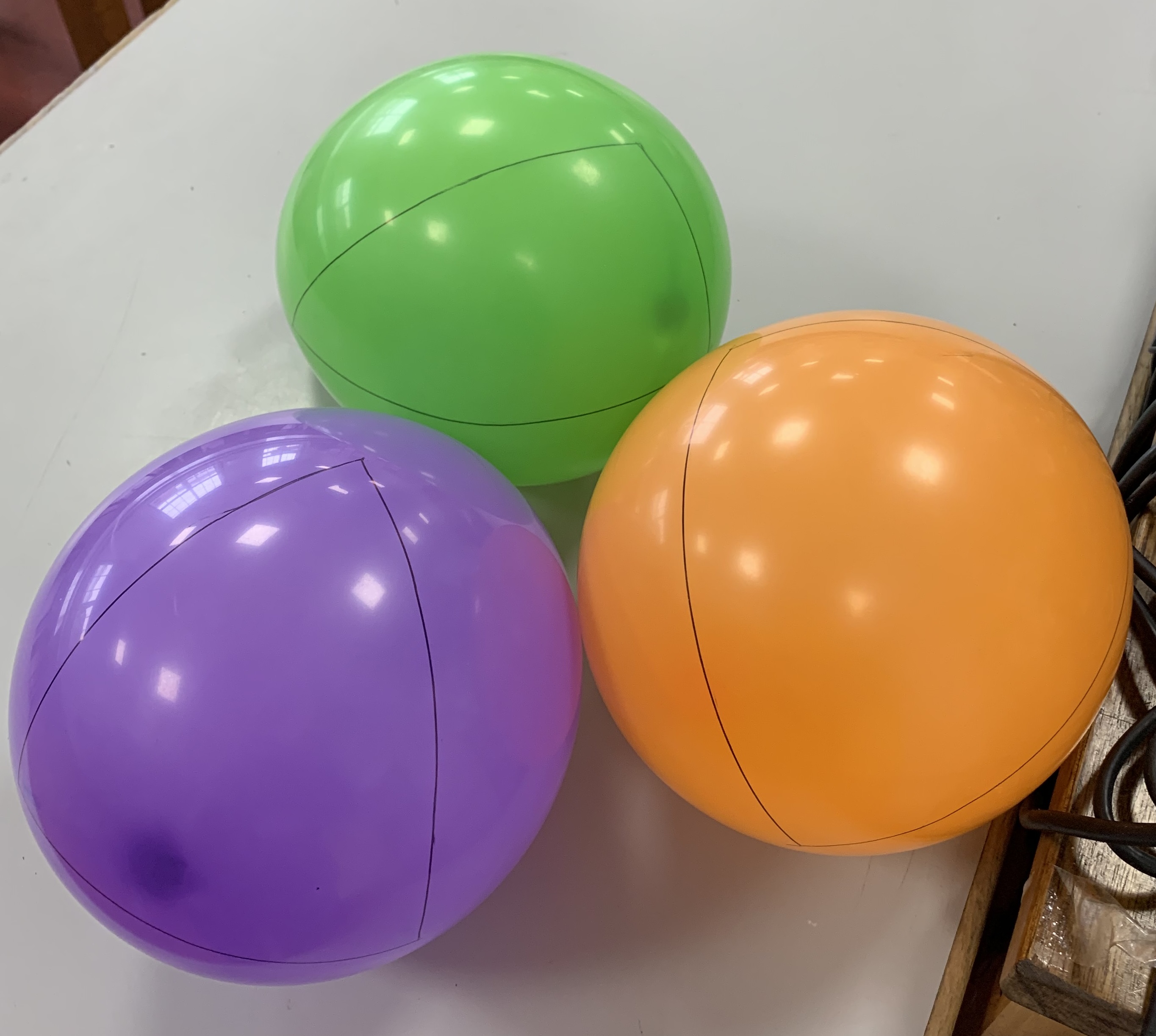}
	\caption{An example of some of the balloons used by students to demonstrate the effects of curvature on geometry}
	\label{fig:Balloons}
\end{figure}

\subsection{Pre/Post Intervention Questionnaire}
\label{Questions}

\begin{table}[h]
\caption{Questions from the pre/post intervention questionnaire}
\label{Table:Questions}
\centerline{%
\begin{tabular}{c|lp{5.5in}|c}
\cline{2-3}
\multicolumn{1}{l|}{}     & \multicolumn{2}{c|}{\textbf{Questions}} & \multicolumn{1}{l}{}        \\ \hline
\multicolumn{1}{|c|}{\multirow{7}{*}{\textbf{Pre}}} & \multicolumn{1}{c|}{\textbf{Q1}}  &     Can parallel lines meet? & \multicolumn{1}{c|}{\multirow{10}{*}{\textbf{Post}}} \\ \cline{2-3}
 \multicolumn{1}{|c|}{}  & \multicolumn{1}{c|}{\textbf{Q2}}  &     Can the sum of the angles in a triangle be different from 180\textsuperscript{$\circ$}?  & \multicolumn{1}{c|}{}  \\
\cline{2-3}
\multicolumn{1}{|c|}{}   &  \multicolumn{1}{c|}{\textbf{Q3}}  &     What is gravity?  & \multicolumn{1}{c|}{}   \\
\cline{2-3}
\multicolumn{1}{|c|}{}  &  \multicolumn{1}{c|}{\textbf{Q4}}  &     How do objects move in gravitational fields?   & \multicolumn{1}{c|}{}  \\
\cline{2-3}
\multicolumn{1}{|c|}{}   & \multicolumn{1}{c|}{\textbf{Q5}}  &      Isaac Newton is famous for his laws of motion and his law of gravity. What is Albert Einstein famous for?  & \multicolumn{1}{c|}{}  \\
\cline{2-3}
\multicolumn{1}{|c|}{}   &  \multicolumn{1}{c|}{\textbf{Q6}}  &     Does space have a shape? What about the space around heavy objects like stars and planets?  & \multicolumn{1}{c|}{}   \\
\cline{2-3}
\multicolumn{1}{|c|}{}   &  \multicolumn{1}{c|}{\textbf{Q7}}  &  Do you prefer to learn about physics by listening to your teacher, watching demonstrations or doing practical work? & \multicolumn{1}{c|}{} \\
\cline{1-3}
 \multicolumn{1}{l|}{} &  \multicolumn{1}{c|}{\textbf{Q8}}  &   Did you enjoy learning about this topic? What did you like/ not like?  & \multicolumn{1}{c|}{}\\
\cline{2-3}
\multicolumn{1}{l|}{}  &  \multicolumn{1}{c|}{\textbf{Q9}}  &   Are you interested in learning more about gravity and general relativity?  & \multicolumn{1}{c|}{}\\
\cline{2-3}
\multicolumn{1}{l|}{}  &  \multicolumn{1}{c|}{\textbf{Q10}} &   Should Einsteinian physics, like general relativity be included in the curriculum? & \multicolumn{1}{c|}{}\\
\cline{2-4}
\end{tabular}%
}
\end{table}

Two sets of questionnaires were used in this research. One disseminated two weeks before the intervention (pre-), and one given to students two weeks after the intervention (post-). Following the format of related studies\textsuperscript{\cite{ChFoKa2018, KaBlBu2018, PiVeBl2014}}, the open-ended questions in the questionnaires examined conceptual understandings and attitudes towards GR.

The pre-questionnaire comprised seven questions exploring students' understanding of gravity, its origins, workings, their perspectives on space shape, and opinions on physics teaching methods. 

The post-questionnaire was nearly identical to the pre-questionnaire, allowing direct comparison. It included three additional questions to gain insight into students' views on the intervention and Einsteinian gravity itself. 

The questions themselves are open-ended and can be found in Table \ref{Table:Questions}.

\section{Results \& Discussion}
\label{Results&Discussion}

\begin{figure}[!ht]
	\centering
	\includegraphics[width=0.9\textwidth,height=0.9\textheight,keepaspectratio]{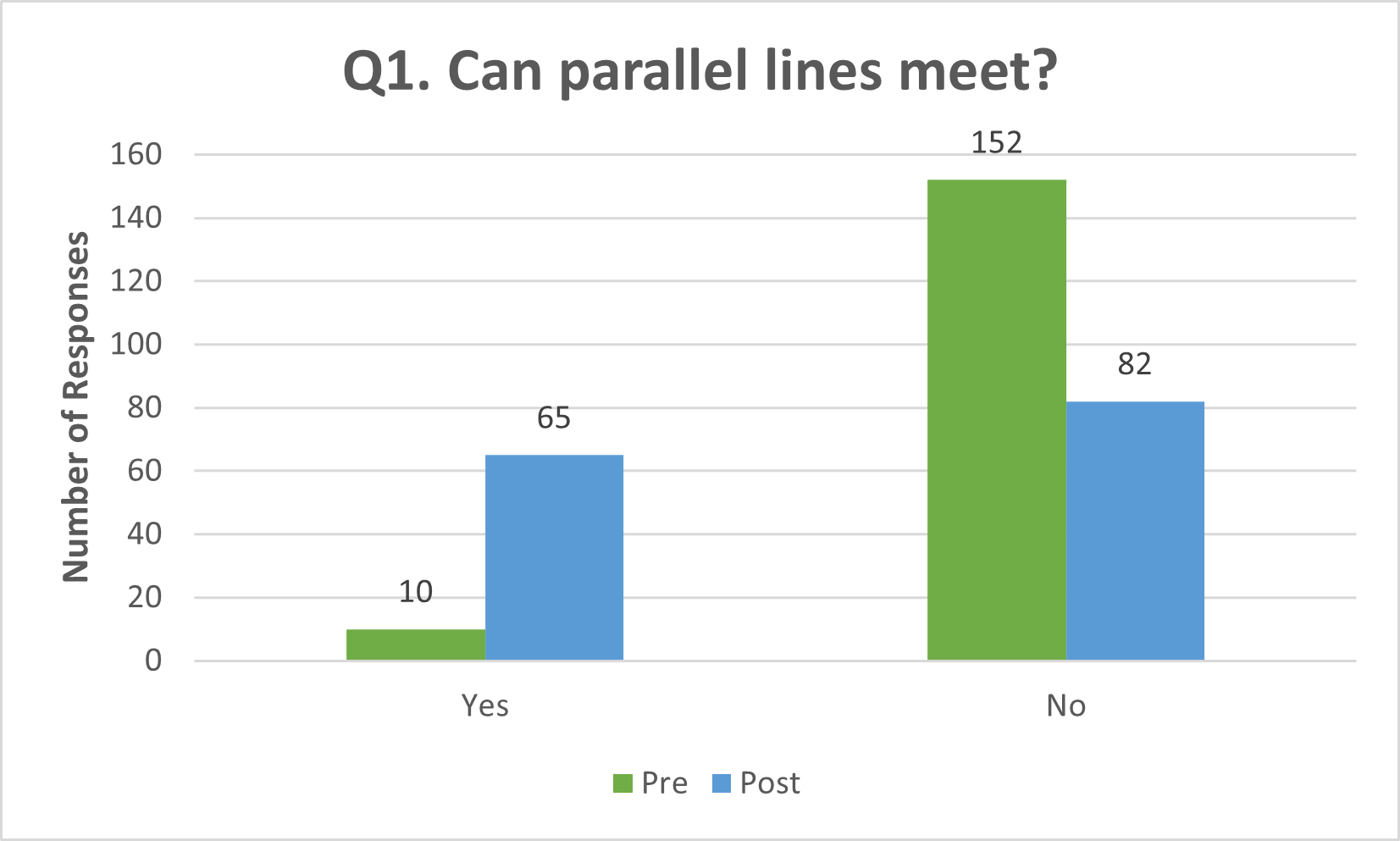}
	\caption{Responses to Q1. Can parallel lines meet?}
	\label{fig:Q1}
\end{figure}

\begin{table}[!ht]
\caption{Difference in pre- and post-questionnaire responses to Q1 by year group.}
\label{Table:Q1}
\centering
\begin{tabular}{l|ll|ll|}
\cline{2-5}
                            & \multicolumn{2}{c|}{Pre}                & \multicolumn{2}{c|}{Post}               \\ \cline{2-5} 
                            & \multicolumn{1}{c|}{No}       & Yes     & \multicolumn{1}{c|}{No}      & Yes      \\ \hline
\multicolumn{1}{|l|}{Yr 7}  & \multicolumn{1}{l|}{93.62\%}  & 8.51\%  & \multicolumn{1}{l|}{79.66\%} & 20.34\%  \\ \hline
\multicolumn{1}{|l|}{Yr 8}  & \multicolumn{1}{l|}{100.00\%} & 0.00\%  & \multicolumn{1}{l|}{75.76\%} & 21.21\%  \\ \hline
\multicolumn{1}{|l|}{Yr 9}  & \multicolumn{1}{l|}{100.00\%} & 0.00\%  & \multicolumn{1}{l|}{33.33\%} & 61.11\%  \\ \hline
\multicolumn{1}{|l|}{Yr 10} & \multicolumn{1}{l|}{78.95\%}  & 21.05\% & \multicolumn{1}{l|}{6.25\%}  & 75.00\%  \\ \hline
\multicolumn{1}{|l|}{Yr 11} & \multicolumn{1}{l|}{91.67\%}  & 8.33\%  & \multicolumn{1}{l|}{20.00\%} & 80.00\%  \\ \hline
\multicolumn{1}{|l|}{Yr 12} & \multicolumn{1}{l|}{91.67\%}  & 8.33\%  & \multicolumn{1}{l|}{0.00\%}  & 100.00\% \\ \hline
\multicolumn{1}{|l|}{Yr 13} & \multicolumn{1}{l|}{90.91\%}  & 9.09\%  & \multicolumn{1}{l|}{11.11\%} & 88.89\%  \\ \hline
\end{tabular}
\end{table}

\begin{figure}[!ht]
	\centering
	\includegraphics[width=0.9\textwidth,height=0.9\textheight,keepaspectratio]{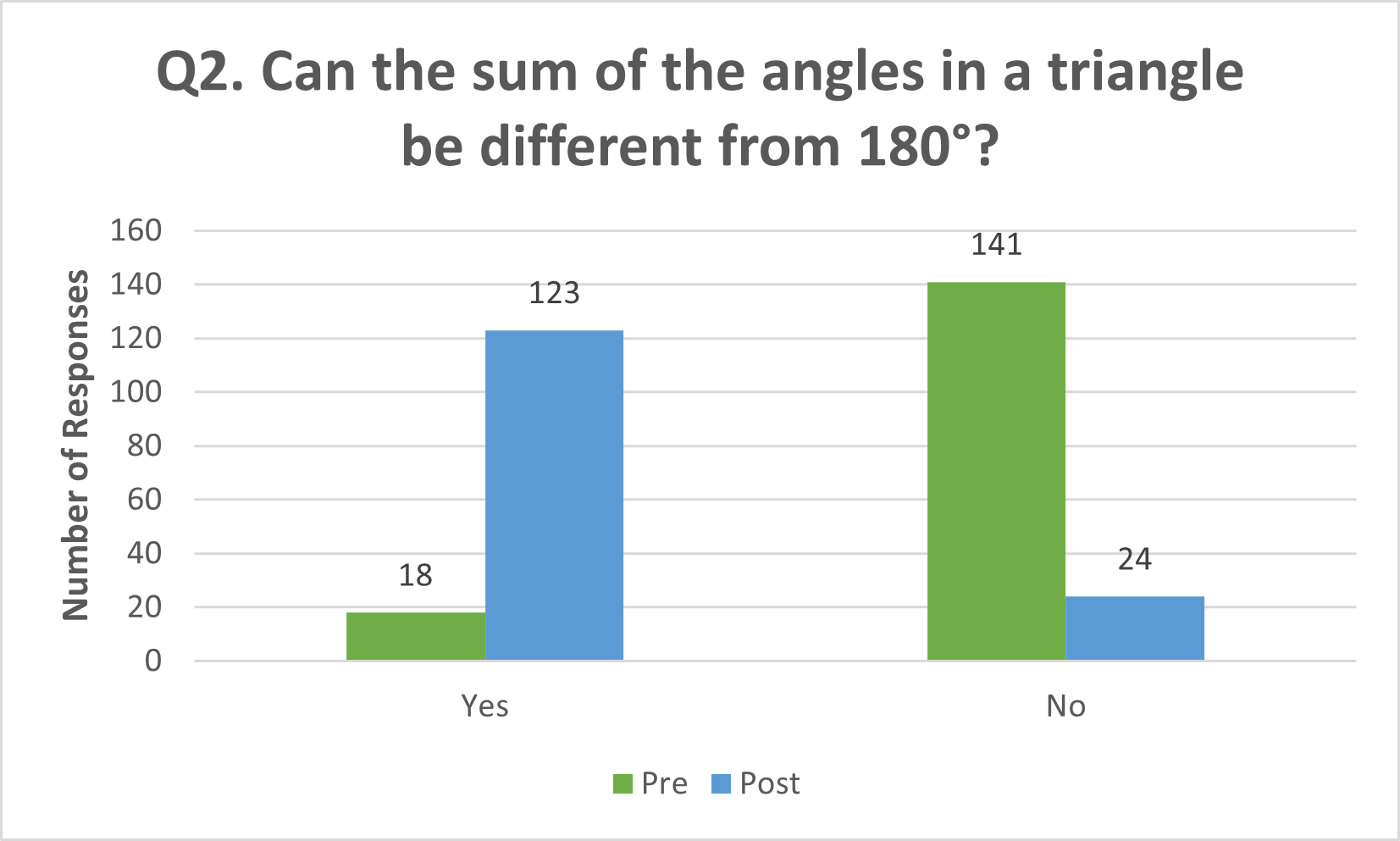}
	\caption{Responses to Q2. Can the sum of the angles in a triangle be different from 180\textsuperscript{$\circ$}?}
	\label{fig:Q2}
\end{figure}

\begin{table}[!ht]
\caption{Difference in pre- and post-questionnaire responses to Q2 by year group.}
\label{Table:Q2}
\centering
\begin{tabular}{l|ll|ll|}
\cline{2-5}
                            & \multicolumn{2}{c|}{Pre}               & \multicolumn{2}{c|}{Post}               \\ \cline{2-5} 
                            & \multicolumn{1}{c|}{No}      & Yes     & \multicolumn{1}{c|}{No}      & Yes      \\ \hline
\multicolumn{1}{|l|}{Yr 7}  & \multicolumn{1}{l|}{93.48\%} & 6.52\%  & \multicolumn{1}{l|}{13.79\%} & 86.21\%  \\ \hline
\multicolumn{1}{|l|}{Yr 8}  & \multicolumn{1}{l|}{88.89\%} & 11.11\% & \multicolumn{1}{l|}{33.33\%} & 63.64\%  \\ \hline
\multicolumn{1}{|l|}{Yr 9}  & \multicolumn{1}{l|}{87.50\%} & 12.50\% & \multicolumn{1}{l|}{16.67\%} & 83.33\%  \\ \hline
\multicolumn{1}{|l|}{Yr 10} & \multicolumn{1}{l|}{84.21\%} & 15.79\% & \multicolumn{1}{l|}{6.25\%}  & 75.00\%  \\ \hline
\multicolumn{1}{|l|}{Yr 11} & \multicolumn{1}{l|}{83.33\%} & 8.33\%  & \multicolumn{1}{l|}{0.00\%}  & 100.00\% \\ \hline
\multicolumn{1}{|l|}{Yr 12} & \multicolumn{1}{l|}{83.33\%} & 16.67\% & \multicolumn{1}{l|}{0.00\%}  & 100.00\% \\ \hline
\multicolumn{1}{|l|}{Yr 13} & \multicolumn{1}{l|}{90.91\%} & 9.09\%  & \multicolumn{1}{l|}{11.11\%} & 88.89\%  \\ \hline
\end{tabular}
\end{table}

\begin{figure}[!ht]

\subfloat[Q3 pre-questionnaire responses\label{Figure5a}]{%
  \includegraphics[clip,width=1.1\columnwidth]{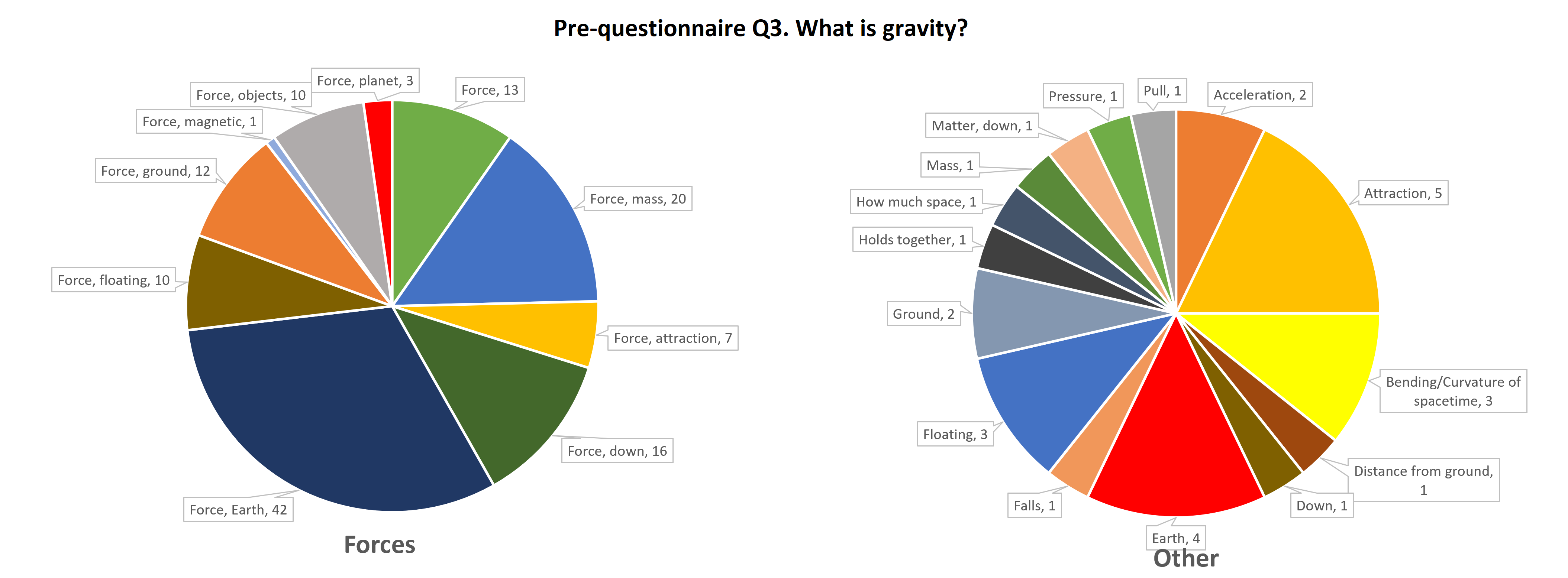}%
}

\subfloat[Q3 post-questionnaire responses\label{Figure5b}]{%
  \includegraphics[clip,width=1.1\columnwidth]{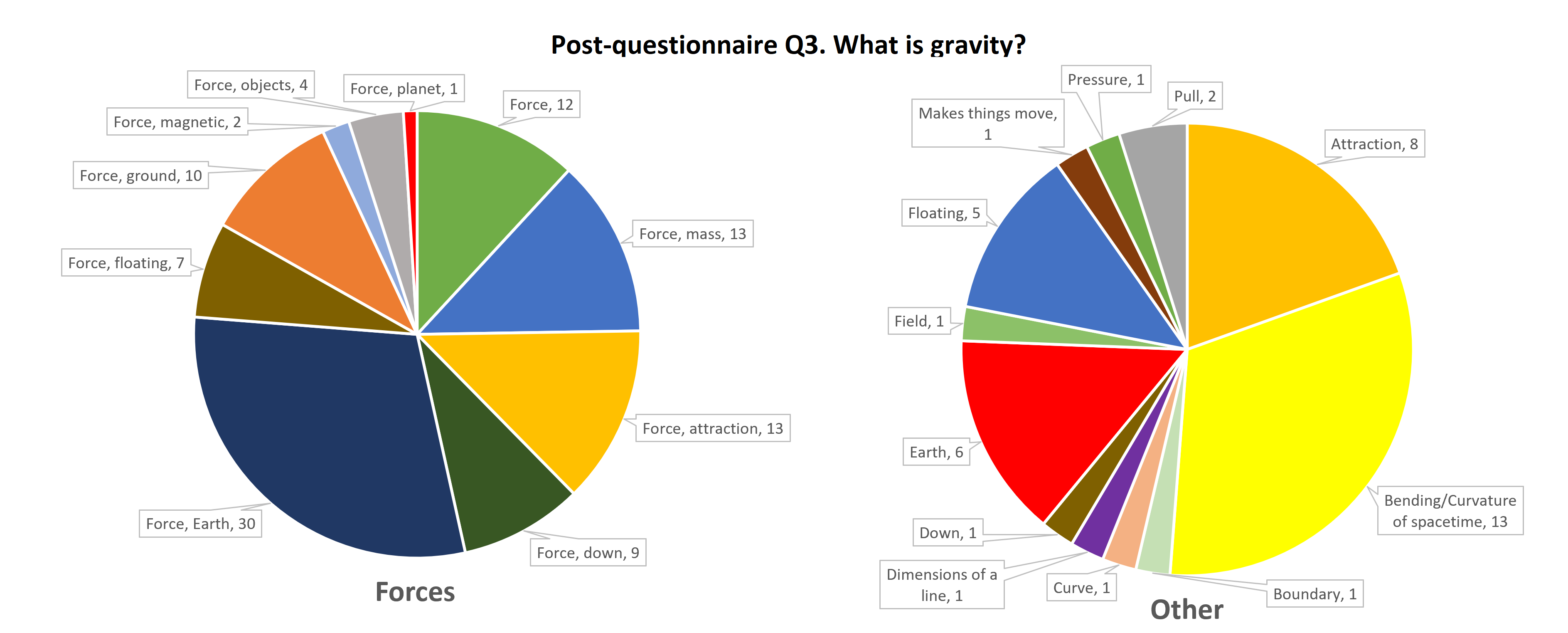}%
}

\caption{Pre- (a) and post-questionnaire (b) responses to Q3. What is gravity? Responses in which gravity is described as a force are shown in the right-hand charts, with other opinions of gravity being shown in the left-hand charts.}
\label{fig:Q3}
\end{figure}

Figure \ref{fig:Q1} and Table \ref{Table:Q1} show the responses to Q1. The majority of participants (93.8\%) said that parallel lines cannot meet. Of the ten that said they can (three from Yr 7, four from Yr 10 and one each from Yr 11, 12 \& 13), five of these responses (three Yr 10, one Yr 11 and one Yr 13) used the phrase `\textit{non-Euclidean}' or `\textit{not on flat space}', thus showing good knowledge before the intervention. It is likely that these participants have had previous exposure to curved geometries. 
The total number of `\textit{yes}' responses moves from 6.2\% to 44.2\% after the intervention. This is not as much as the 11.3\% to 83.6\% rise in correct responses to Q2, shown in Figure \ref{fig:Q2} and Table \ref{Table:Q2}. The intervention design contributed to this as Q2 relates to the balloon activity, whereas the information about parallel lines also being influenced by curved geometry was presented after this activity as part of the slideshow presentation. As such, participants had to pay more attention to grasp this information. Nonetheless, all year groups showed an increase in `\textit{yes}' responses for both Q1 and Q2 with Yr 9, 11, and 12 having the highest increase in percentage. This is perhaps expected as these cohorts have studied more geometry than younger students and so are better prepared for discussion's around non-Euclidean geometry.

These results indicate that participants comprehend information better through visual and hands-on activities. Well-designed practical activities are reported to increase students levels of understanding\textsuperscript{\cite{AbMi2008}}, with practical work seen by students as more interesting and engaging than listening to their teacher or watching demonstrations\textsuperscript{\cite{Millar2010}}.

\begin{table}[!ht]
\caption{Difference in pre- and post-questionnaire responses to Q3 by year group.}
\label{Table:Q3}
\centerline{%
\begin{tabular}{l|llll|llll|}
\cline{2-9}
                            & \multicolumn{4}{c|}{Pre}                                                                                                                                              & \multicolumn{4}{c|}{Post}                                                                                                                                             \\ \cline{2-9} 
                            & \multicolumn{2}{c|}{Force}                                  & \multicolumn{1}{c|}{\multirow{2}{2cm}{Spacetime Curvature}} & \multicolumn{1}{c|}{\multirow{2}{*}{Earth}} & \multicolumn{2}{c|}{Force}                                  & \multicolumn{1}{c|}{\multirow{2}{2cm}{Spacetime Curvature}} & \multicolumn{1}{c|}{\multirow{2}{*}{Earth}} \\ \cline{2-3} \cline{6-7}
                            & \multicolumn{1}{l|}{Earth}   & \multicolumn{1}{l|}{Other}   & \multicolumn{1}{c|}{}                                     & \multicolumn{1}{c|}{}                       & \multicolumn{1}{l|}{Earth}   & \multicolumn{1}{l|}{Other}   & \multicolumn{1}{c|}{}                                     & \multicolumn{1}{c|}{}                       \\ \hline
\multicolumn{1}{|l|}{Yr 7}  & \multicolumn{1}{l|}{25.53\%} & \multicolumn{1}{l|}{48.94\%} & \multicolumn{1}{l|}{0.00\%}                               & 6.38\%                                      & \multicolumn{1}{l|}{17.24\%} & \multicolumn{1}{l|}{48.28\%} & \multicolumn{1}{l|}{1.72\%}                               & 6.90\%                                      \\ \hline
\multicolumn{1}{|l|}{Yr 8}  & \multicolumn{1}{l|}{35.14\%} & \multicolumn{1}{l|}{54.05\%} & \multicolumn{1}{l|}{0.00\%}                               & 2.70\%                                      & \multicolumn{1}{l|}{25.81\%} & \multicolumn{1}{l|}{51.61\%} & \multicolumn{1}{l|}{3.23\%}                               & 6.45\%                                      \\ \hline
\multicolumn{1}{|l|}{Yr 9}  & \multicolumn{1}{l|}{29.17\%} & \multicolumn{1}{l|}{87.50\%} & \multicolumn{1}{l|}{0.00\%}                               & 0.00\%                                      & \multicolumn{1}{l|}{38.89\%} & \multicolumn{1}{l|}{38.89\%} & \multicolumn{1}{l|}{5.56\%}                               & 0.00\%                                      \\ \hline
\multicolumn{1}{|l|}{Yr 10} & \multicolumn{1}{l|}{5.56\%}  & \multicolumn{1}{l|}{72.22\%} & \multicolumn{1}{l|}{5.56\%}                               & 0.00\%                                      & \multicolumn{1}{l|}{15.38\%} & \multicolumn{1}{l|}{53.85\%} & \multicolumn{1}{l|}{23.08\%}                              & 0.00\%                                      \\ \hline
\multicolumn{1}{|l|}{Yr 11} & \multicolumn{1}{l|}{58.33\%} & \multicolumn{1}{l|}{41.67\%} & \multicolumn{1}{l|}{0.00\%}                               & 0.00\%                                      & \multicolumn{1}{l|}{30.00\%} & \multicolumn{1}{l|}{40.00\%} & \multicolumn{1}{l|}{10.00\%}                              & 0.00\%                                      \\ \hline
\multicolumn{1}{|l|}{Yr 12} & \multicolumn{1}{l|}{16.67\%} & \multicolumn{1}{l|}{75.00\%} & \multicolumn{1}{l|}{8.33\%}                               & 0.00\%                                      & \multicolumn{1}{l|}{0.00\%}  & \multicolumn{1}{l|}{42.86\%} & \multicolumn{1}{l|}{28.57\%}                              & 0.00\%                                      \\ \hline
\multicolumn{1}{|l|}{Yr 13} & \multicolumn{1}{l|}{0.00\%}  & \multicolumn{1}{l|}{81.82\%} & \multicolumn{1}{l|}{9.09\%}                               & 0.00\%                                      & \multicolumn{1}{l|}{0.00\%}  & \multicolumn{1}{l|}{44.44\%} & \multicolumn{1}{l|}{44.44\%}                              & 0.00\%                                      \\ \hline
\end{tabular}%
}
\end{table}

Q3 is more open to predisposed misconceptions than Q1 or Q2. As seen in Figure \ref{fig:Q3}, there were many different responses to this question. We have categorised responses into those which described gravity as a force versus those which did not. It was expected that post-intervention, participants would shift their answers towards describing gravity as the `\textit{bending/warping of spacetime}'. 82\% of the pre-questionnaire responses described gravity as a force, with responses ranging from `\textit{the force that keeps us on Earth}', `\textit{the force that stops us from floating/keeps us on the ground}' and `\textit{the force between objects/masses}'.

Only 1.8\% of participants specified that gravity was the cause of spacetime curvature pre-intervention, increasing to only 8.9\% post-intervention. Similar studies\textsuperscript{\cite{PoDe2021}} also struggled to fully influence descriptions of gravity using the spacetime simulator. However looking at Table \ref{Table:Q3} which shows the breakdown of Q3 by year group, reveals that responses related to spacetime curvature increased notably in years 10 (up 17.52\%), 12 (up 20.24\%), and 13 (up 35.35\%).

One positive result from Table \ref{Table:Q3} is that participants associating gravity as an Earth-bound force decreased across all year groups (except Yr 10), particularly in Yr 8 and 11. The data indicates that the KS4 and A-Level groups best understood the concept of gravity from spacetime curvature. The majority of participants however either did not grasp the concepts introduced (not indicated by the results to Q1 and Q2), or the intervention's influence was limited by their pre-existing classroom-based learning and opinions.

\begin{landscape}
    \begin{figure}[!ht]
    	\centering
    	\includegraphics[width=1.3\textwidth, height=\textheight]{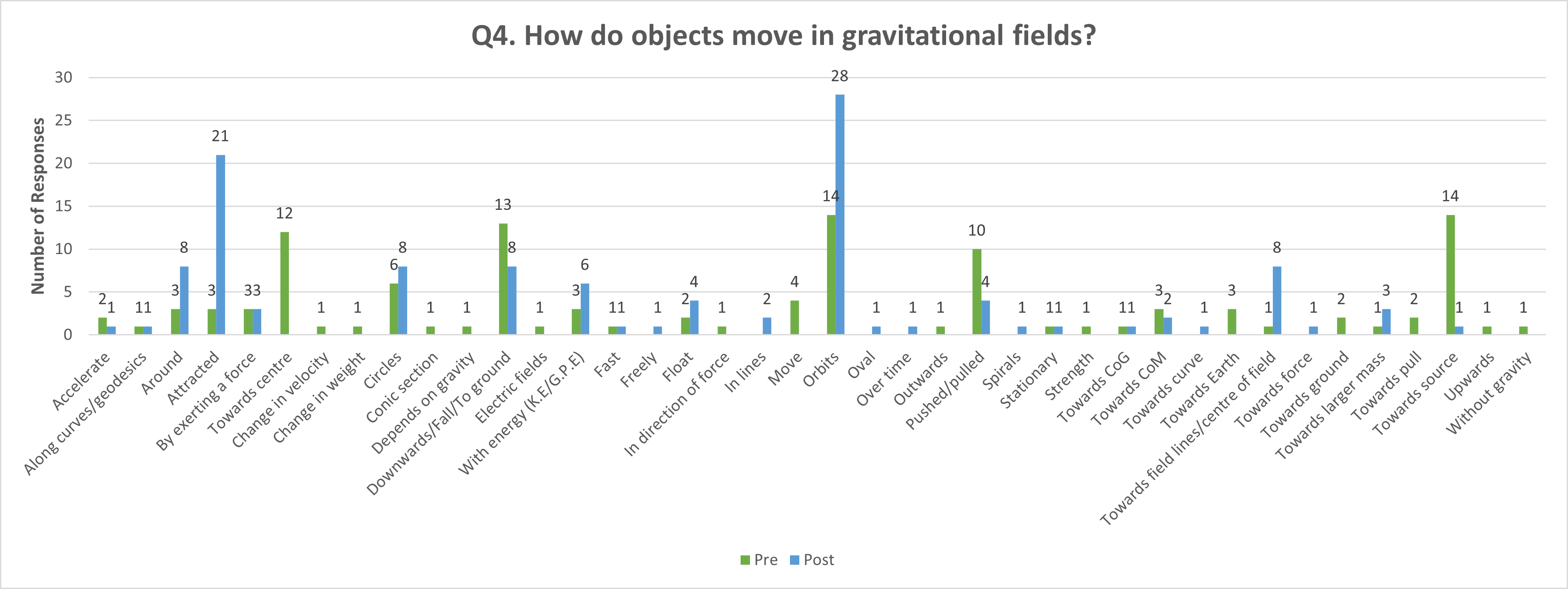}
    	\caption{Responses to Q4. How do objects move in gravitational fields?}
    	\label{fig:Q4}
    \end{figure}
\end{landscape}

Q4 assess participants understanding of how gravity influences the motion of objects. Presented in Figure \ref{fig:Q4}, the responses to this question are extremely varied, ranging from high level answers like `\textit{conic sections}' and `\textit{along geodesics}' to less descriptive responses such as `\textit{they float}' or `\textit{fast}'. Many responses show an understanding of the attractive powers of gravity, as well as knowledge of other aspects of physics such as kinetic and gravitational potential energy. While many of these descriptions are not wholly incorrect, they fail to properly describe the motion of objects in gravitational fields. This is likely due to participants lack of formal education on gravitational orbits, a topic not covered in detail until Yr 13. It is possible that responses such as `\textit{around}' and `\textit{in circles}' are consequences of this as those in lower year groups attempt to describe orbits without knowledge of the correct scientific terminology. This is supported by the fact that we see a slight increase in these descriptors post-intervention as participants watched objects circle around masses on the spacetime simulator sheet. 

Positively, phrases such as `\textit{they fall}', `\textit{towards ground/source}' and `\textit{pushed/pulled}' decrease post-intervention, whereas the number of participants describing orbital motion doubles. Those describing motion in gravitational fields as attractive also sees a significant increase post-intervention. This again is likely due to the use of the spacetime simulator demonstration where it was observed that objects on the sheet moved towards the source of curvature. This is also evidenced by the added uses of terms such as '\textit{oval}' and '\textit{spirals}' which appear in post-questionnaire responses only.

\begin{figure}[!ht]
	\centering
	\includegraphics[width=1.1\textwidth,height=1.1\textheight,keepaspectratio]{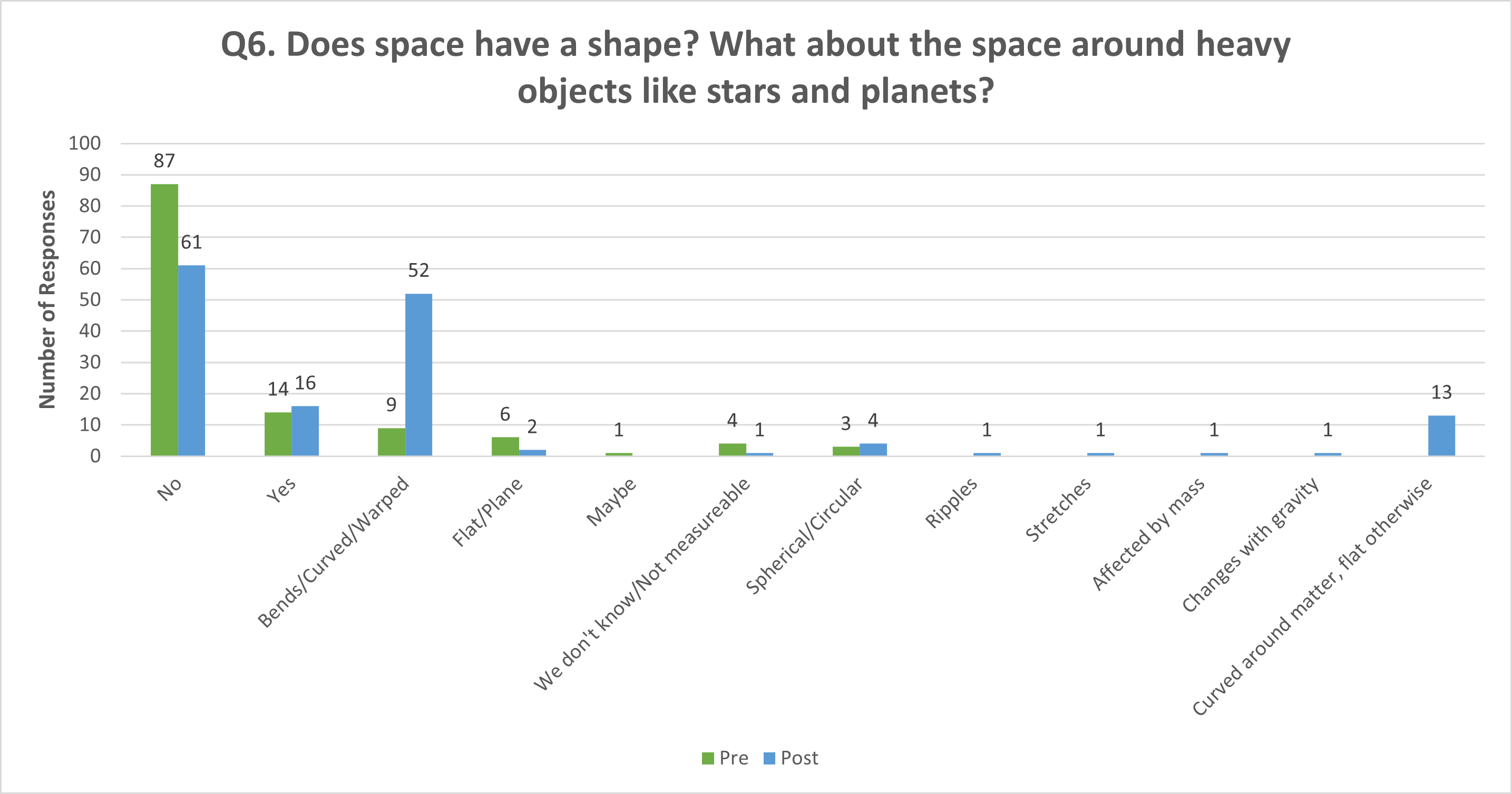}
	\caption{Responses to Q6. Does space have a shape? What about the space around heavy objects like stars and planets?}
	\label{fig:Q6}
\end{figure}

Pre-intervention, 58.2\% of participants thought that space had no shape, often saying it is infinite or expanding (`\textit{no because it is infinite/expanding}') to justify their response. This number decreases to 42.4\% post-intervention, indicating that the intervention had limited success in conveying the geometry of space around celestial objects. Nonetheless, Figure \ref{fig:Q6} reveals an increase in responses other than `\textit{yes}' or `\textit{no}' post-intervention with the addition of more conceptualised answers like `\textit{the shape changes with gravity}' or `\textit{it stretches}' demonstrating an understanding of the ideas presented using the spacetime simulator. 9.2\% of post-intervention participants shown deeper understanding by correctly noting that space is curved around masses but flat otherwise. Coupling these responses with the other `\textit{yes}' responses gives a total percentage of conceptually correct answers of 58.9\%. 

\begin{figure}[!ht]
	\centering
	\includegraphics[width=\textwidth,height=\textheight,keepaspectratio]{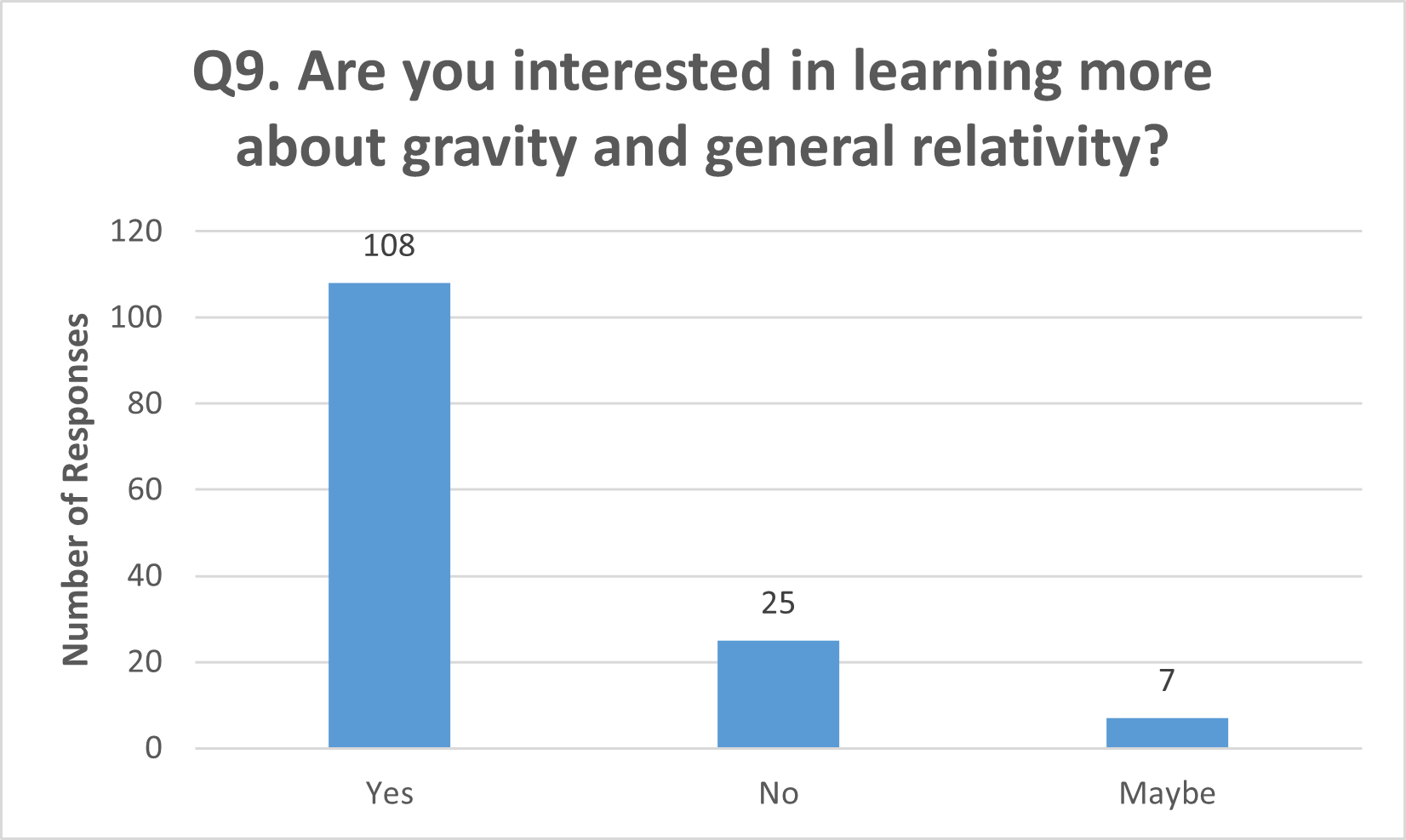}
	\caption{Responses to Q9. Are you interested in learning more about gravity and general relativity?}
	\label{fig:Q9}
\end{figure}

Figure \ref{fig:Q9} shows responses to Q9, where 77.1\% of participants said they would like to learn more about gravity and general relativity.

\begin{figure}[!ht]
	\centering
	\includegraphics[width=\textwidth,height=\textheight,keepaspectratio]{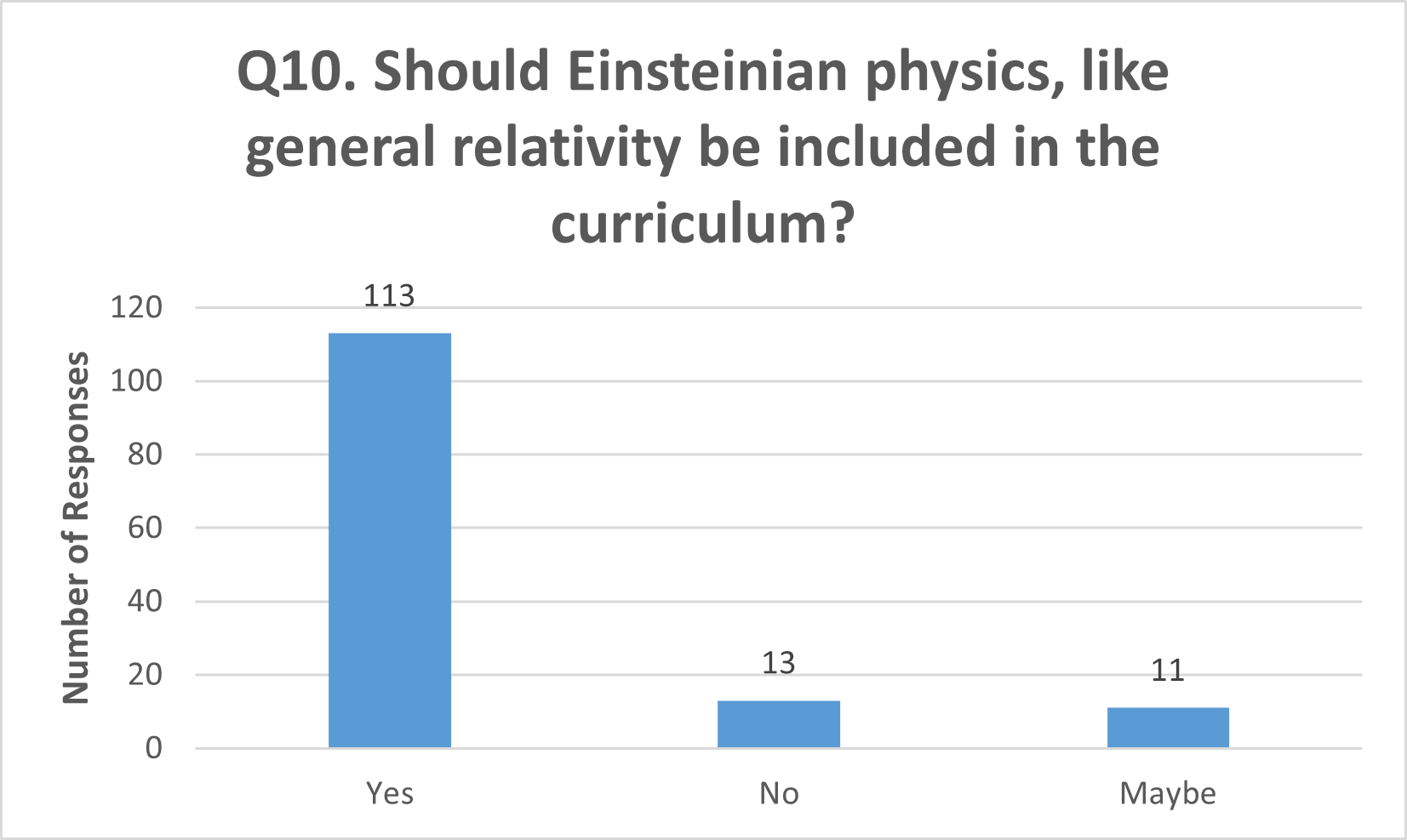}
	\caption{Responses to Q10. Should Einsteinian physics, like general relativity be included in the curriculum?}
	\label{fig:Q10}
\end{figure}

The last question in our questionnaires addresses the main research question of this work and asked 'Should Einsteinian physics, like general relativity be included in the curriculum? A resounding 82.5\% of participants said that it should be included. Of those, 10\% indicated that, if added to the curriculum, it should be included at a specific level. 41\% of those respondents said that this content should not be taught before A-Level. The rest thought that it should not be covered before KS4.

\section{Conclusion}
\label{Conclusion}

In this work, 183 students from a state-funded English secondary school participated in a one-off intervention introducing them to General Relativity and the concept that the gravitational force is caused by the curvature of spacetime. This theory was introduced visually using the spacetime simulator demonstration. Once this idea was established, participants performed an experiment using balloons and triangles to investigate what happens to these shapes in curved spaces.

The aim was to gauge students' reception to GR and to investigate where these concepts would be best placed into a school curriculum. 

Results showed that 77.1\% of our participants overall would like to delve further into the subject. This is consistent with the opinions of teachers and the public who also encourage the addition of Einsteinian physics into the curriculum\textsuperscript{\cite{FoChBl2019, BaErOl2022}}. Noting that the largest percentage of our participants come from Yr 7 shows how eager young students are for opportunities to learn about cosmology.

KS4 and A-Level students demonstrated better understanding of the presented ideas through their responses to Q1-3 in the questionnaires. Years 11, 12 and 13 had the best improvement from pre- to post-questionnaire for Q1 and Q2, with Yr 10 also showing a good increase in understanding for Q3. While many participants from the other year groups liked the material covered, their responses to Q3 showed little grasp of spacetime curvature. Years 7, 8 and 9 all showed good awareness of the deformation of triangles in curved spaces after the intervention, only Yr 9 participants showed a grasp of the curving of initially parallel lines.  

While Einsteinian gravitational concepts are taught completely distinct from the Newtonian gravity, it has been shown by other studies that students can understand the ideas of GR even at primary level\textsuperscript{\cite{RuMaLe2021, KaBlMo2017b, PiVeBl2014, Elise2007}}. While the results to Q1 and Q2 show that participants were able to appreciate the effects of curvature, the results of Q3 demonstrate that it would take more than a one-off intervention to imbue a deeper understanding of GR. This result aligns with that of other works\textsuperscript{\cite{ChKrKe2019}}. Additionally, our intervention was not successful at dispelling misconceptions around the gravitational force. A multi-stage intervention may prove more beneficial for this\textsuperscript{\cite{ChFoKa2022}}. 

Regarding GR and its associated geometry, our results show that this material is better suited to the A-Level curriculum rather than lower year groups. Therefore, we conclude that introducing GR in the English secondary curriculum at Yr 13 aligns with students' ability to recognize curvature as the source of gravitational force and matches the current placement of gravity study in the curriculum.

Further insights into the effects of teaching GR could be obtained through a longitudinal study that follows a group or groups of students through late primary/early secondary school to KS4 and A-Level.

\begin{description}
    \item[\textbf{Ethical Statement:}] This study has been reviewed and given favourable opinion by a University of Lincoln Research Ethics Committee. Reference Number: 12146.    
\end{description}

\ack

The authors would like to thank the staff and students of the science department at Queen Elizabeth's High School for their support with and participation in this work.


\printbibliography

\appendix

\begin{appendices}
\section{Supplimentary Results}

\begin{table}[!ht]
\caption{Number of questionnaire responses}
\label{Table:Responses}
\begin{subtable}[c]{0.5\textwidth}
\centering
\begin{tabular}{lll}
\cline{2-3}
\multicolumn{1}{l|}{}    & \multicolumn{1}{l|}{Pre} & \multicolumn{1}{l|}{Post} \\ \hline
\multicolumn{1}{|l|}{Yr 7}  & \multicolumn{1}{l|}{47}  & \multicolumn{1}{l|}{59}   \\ \hline
\multicolumn{1}{|l|}{Yr 8}  & \multicolumn{1}{l|}{37}  & \multicolumn{1}{l|}{33}   \\ \hline
\multicolumn{1}{|l|}{Yr 9}  & \multicolumn{1}{l|}{24}  & \multicolumn{1}{l|}{18}   \\ \hline
\multicolumn{1}{|l|}{Yr 10} & \multicolumn{1}{l|}{19}  & \multicolumn{1}{l|}{16}   \\ \hline
\multicolumn{1}{|l|}{Yr 11} & \multicolumn{1}{l|}{12}  & \multicolumn{1}{l|}{10}   \\ \hline
\multicolumn{1}{|l|}{Yr 12} & \multicolumn{1}{l|}{12}  & \multicolumn{1}{l|}{7}    \\ \hline
\multicolumn{1}{|l|}{Yr 13} & \multicolumn{1}{l|}{11}  & \multicolumn{1}{l|}{9}    \\ \hline
                         &                          &                           \\
                         &                          &                           \\
                         &                          &                          
\end{tabular}
\subcaption{Number of responses by year group}
\end{subtable}
\begin{subtable}[c]{0.5\textwidth}
\centering
\begin{tabular}{l|ll|}
\cline{2-3}
                          & \multicolumn{1}{l|}{Pre}       & Post      \\ \hline
\multicolumn{1}{|l|}{Q1}  & \multicolumn{1}{l|}{162}       & 147       \\ \hline
\multicolumn{1}{|l|}{Q2}  & \multicolumn{1}{l|}{161}       & 147       \\ \hline
\multicolumn{1}{|l|}{Q3}  & \multicolumn{1}{l|}{161}       & 146       \\ \hline
\multicolumn{1}{|l|}{Q4}  & \multicolumn{1}{l|}{131}       & 124       \\ \hline
\multicolumn{1}{|l|}{Q5}  & \multicolumn{1}{l|}{147}       & 140       \\ \hline
\multicolumn{1}{|l|}{Q6}  & \multicolumn{1}{l|}{139}       & 141       \\ \hline
\multicolumn{1}{|l|}{Q7}  & \multicolumn{1}{l|}{150}       & 144       \\ \hline
\multicolumn{1}{|l|}{Q8}  & \multicolumn{1}{l|}{}          & 138       \\ \cline{1-1} \cline{3-3} 
\multicolumn{1}{|l|}{Q9}  & \multicolumn{1}{l|}{}          & 143       \\ \cline{1-1} \cline{3-3} 
\multicolumn{1}{|l|}{Q10} & \multicolumn{1}{l|}{}          & 141       \\ \cline{1-1} \cline{3-3} 
\end{tabular}
\subcaption{Number of responses by question}
\end{subtable}
\end{table}

\begin{figure}[!ht]

\subfloat[Q5 pre-questionnaire responses\label{fig:Q5Pre}]{%
  \includegraphics[clip,width=1.1\columnwidth]{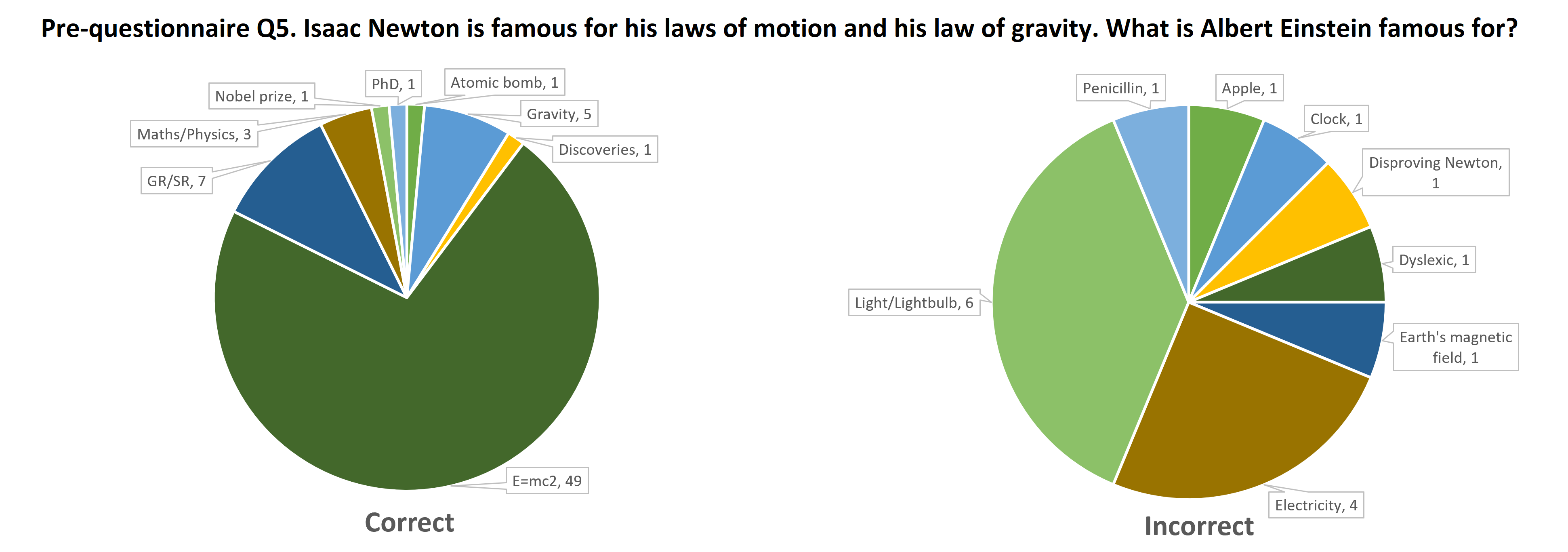}%
}

\subfloat[Q5 post-questionnaire responses\label{fig:Q5Post}]{%
  \includegraphics[clip,width=1.1\columnwidth]{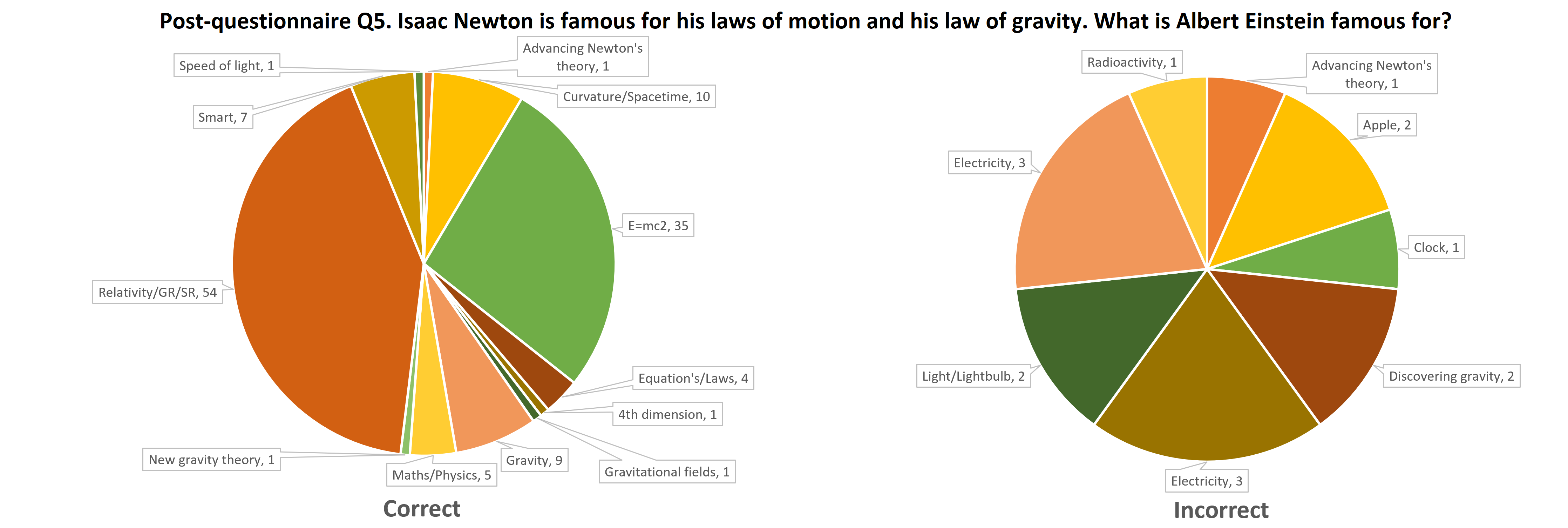}%
}

\caption{Pre- (a) and post-questionnaire (b) responses to Q5. Isaac Newton is famous for his laws of motion and his law of gravity. What is Albert Einstein famous for? Responses are loosely categorised into correct (right-hand charts) and incorrect (left-hand charts) responses.}
\label{fig:Q5}
\end{figure}

It is recognised that teaching the history of physics can be beneficial in increasing students understanding of material\textsuperscript{\cite{Seroglou2001, Galili2008}}. Q5 in our study was designed to probe students background knowledge of one of the most famous faces in all of science and the person behind the theories presented during our intervention, Albert Einstein. Easily recognisable, with his large white hair and distinctive mustache, Einstein's image forms that of the stereotypical `mad scientist'\textsuperscript{\cite{Rowe2012}} and is likely to have been seen in some capacity by many of our participants. But, how many of those familiar with what the man looks like are also familiar with his works? That is the goal of Q5, the results of which are presented in Figure \ref{fig:Q5}. Here, questionnaire responses are loosely sectioned into correct ($E=mc^2$, General and Special Relativity etc.) and incorrect answers (light-bulb, electricity etc.). 

Even pre-intervention, it can be seen that many participants were already familiar with Einstein's famous equation, $E=mc^2$, with a small amount also familiar with the theory of relativity. It must be noted however, that during the intervention few participants were able to define the terms of this equation or knew what it related to.

62.5\% of those with incorrect responses associated Einstein with the light-bulb or electricity. It is difficult to know where these misconceptions originated from. An argument could be made for Einstein's discovery of the photoelectric effect or his use of the speed of light in $E=mc^2$, but these details were not present in the responses of these participants.

Post-intervention, correct responses about Einstein increase from  48.6\% to 90.7\%. Although some participants have missed the mark with responses such as '\textit{advancing Newton's theory}' or '\textit{discovering gravity}'.

\begin{figure}[!ht]
	\centering
	\includegraphics[width=1.1\textwidth,height=1.1\textheight,keepaspectratio]{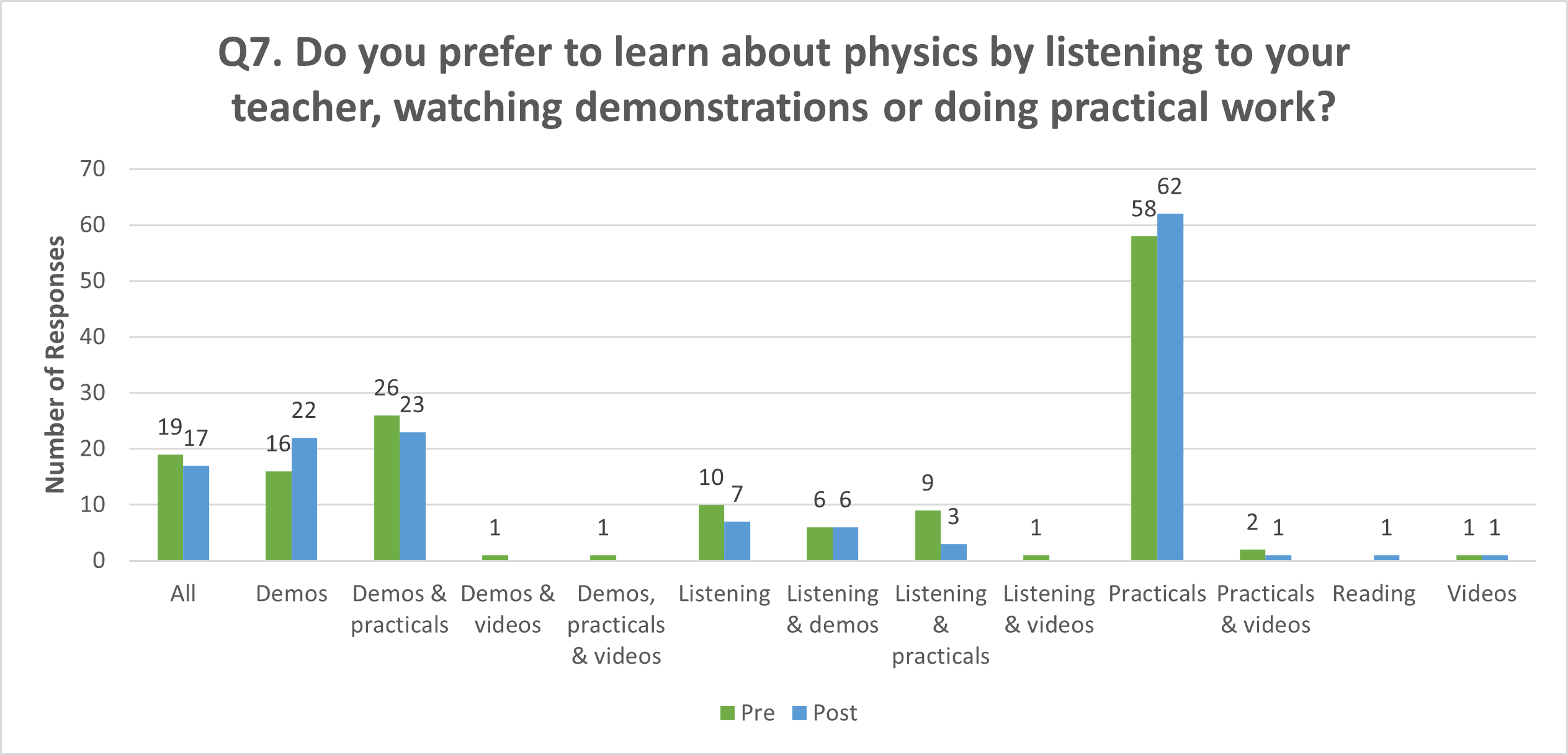}
	\caption{Responses to Q7. Do you prefer to learn about physics by listening to your teacher, watching demonstrations or doing practical work?}
	\label{fig:Q7}
\end{figure}

Q7 of our questionnaire, shown in Figure \ref{fig:Q7}, asked participants about their preferred method of learning physics and was included in both pre- and post-questionnaires for consistency only. It was not expected that there would be much difference in responses between questionnaires considering the short time between them. A large majority of our participants enjoy learning physics via practical work with the next most popular response being a mixture of teacher demonstrations and practical activities. This was a consistent theme throughout the year groups. Interestingly, the use of alternative techniques such as thought-experiments\textsuperscript{\cite{BrBr2012, VeHa2013}} and augmented reality technology\textsuperscript{\cite{ViSaMe2021}} have been successful in engaging students with gravity and GR lessons.

\begin{figure}[!ht]
	\centering
	\includegraphics[width=\textwidth,height=\textheight,keepaspectratio]{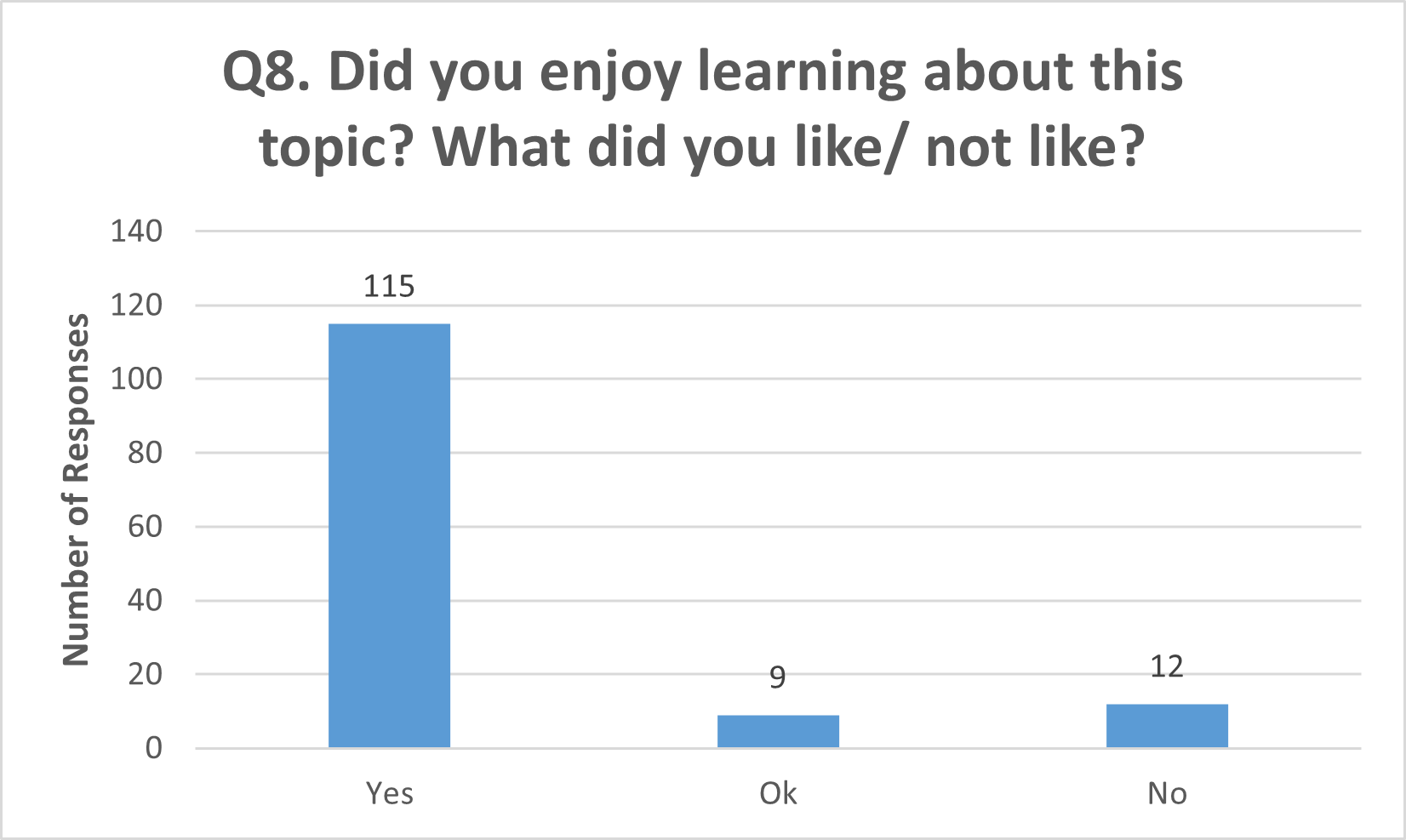}
	\caption{Responses to Q8. Did you enjoy learning about this topic? What did you like/ not like?}
	\label{fig:Q8}
\end{figure}

Shown in Figure \ref{fig:Q8}, 84.6\% of participants said they enjoyed learning about the curvature of spacetime and curved geometries. Of those responses to Q8 which indicated which aspects of the intervention they enjoyed, 50\% enjoyed the spacetime simulator most, 36.4\% enjoyed the practical with the balloons with the rest enjoying being introduced to new and `\textit{interesting}' material. 

\end{appendices}

\end{document}